\documentclass[twocolumn,secnumarabic,amssymb, bibnotes, showpacs, superscriptaddress, aps, prl]{revtex4-1}

\usepackage[inline]{enumitem}

\usepackage{todonotes}                                   

\usepackage{bm}                                              
\usepackage{amssymb}   				
\usepackage{mathtools} 			           
\usepackage[version=3]{mhchem}                   

\usepackage{natbib}

\begin{document}

\title{Fast data sorting with modified principal component analysis to distinguish unique single molecular break junction trajectories}%

\author{J.M.~Hamill} \affiliation{Department of Chemistry and Biochemistry, University of Bern, Freiestrasse 3, CH-3012, Bern, Switzerland}
\author{X.T.~Zhao} \affiliation{Department of Chemistry, Durham University, Durham DH1 3LE, United Kingdom}
\author{G.~M\'{e}sz\'{a}ros} \affiliation{Research Centre for Natural Sciences, Hungarian Academy of Sciences, Magyar Tud\'{o}sok K\"{o}r\'{u}tja 2, H-1117 Budapest, Hungary}
\author{M.R.~Bryce} \affiliation{Department of Chemistry, Durham University, Durham DH1 3LE, United Kingdom}
\author{M.~Arenz} \affiliation{Department of Chemistry and Biochemistry, University of Bern, Freiestrasse 3, CH-3012, Bern, Switzerland}
\email[Author email: ]{joseph.hamill@dcb.unibe.ch}
\date{\today}%

\begin{abstract}
A simple and fast analysis method to sort large data sets into groups with shared distinguishing characteristics is described, and applied to single molecular break junction conductance versus electrode displacement data. The method, based on principal component analysis, successfully sorted data sets based on the projection of the data onto the first or second principal component of the correlation matrix without the need to assert any specific hypothesis about the expected features within the data. This was an improvement on the current correlation matrix analysis approach because it sorted data automatically, making it more objective and less time consuming, and our method is applicable to a wide range of multivariate data sets. Here the method was demonstrated on two systems. First, it was demonstrated on mixtures of two molecules with identical anchor groups, similar lengths, but either a $\pi$ (high conductance) or $\sigma$ (low conductance) bridge. The mixed data was automatically sorted into two groups containing one molecule or the other. Second, it was demonstrated on break junction data measured with the $\pi$ bridged molecule alone. Again the method distinguished between two groups. These groups were tentatively assigned to different geometries of the molecule in the junction.
\end{abstract}


\maketitle

Extremely large data sets, especially as a result of automated measurements, are becoming more common. Quick and powerful methods are required to sift through the data and organize it to save room on hard drives, save time in analysis, and focus attention on significant results. This must be done in unbiased and rigorous ways. Here data sets of single molecular break junction (SMBJ) conductance traces were measured and a new statistical analysis method was used to sort the data into groups with unique characteristics. We first demonstrate that this method can separate the distinctive junctions of one molecule or the other when the molecules are mixed in solution. We then demonstrate that the same method can sort junctions with one single molecule alone into separate groups of junctions. This allowed for a more comprehensive analysis of the experimental results.

Automated SMBJ experiments result in terabytes of traces with a broad sampling of gold-molecule-gold junctions including the following types: 
\begin{enumerate*}
\item junctions which do not break cleanly due to contamination or environmental disruptions;
\item junctions which break cleanly but contain no molecule;
\item junctions which break cleanly and contain multiple molecules;
\item and junctions that break cleanly and contain a single molecule.
\end{enumerate*} 
Usually only junctions of type 4 are  interesting for further analysis. Pre-analysis methods are often employed to filter the data, with varying degrees of objectivity. Recently new methods\cite{Magyarkuti2017, Lemmer2016, Inkpen2015} analyze large SMBJ data sets without making \textit{a priori} assumptions about molecular plateau shape. 

Type 4 junction trajectories have a wide degree of variability due to a number of stochastic processes involved\cite{Guo2011, Malen2009}. Calculations predict correlations between junction geometries and molecular conductance.\cite{Kamenetska2009} However, conductance changes between different junction geometries are often indistinguishable. 

The method developed by Halbritter \textit{et al}\cite{Halbritter2010} distinguishes important junction trajectories. The method calculates a 1D histogram from each conductance versus displacement trace, and compiles them into a data matrix, $\mathbf{X}$, with $m$ rows of conductance bins and $n$ separate measurements.The correlation matrix, $C$, for $\mathbf{X}$ is calculated by
\begin{equation}\label{eq:2DCH}
C_{i,j}=\frac
{\langle \left[ x_i - \langle x_i \rangle \right] * \left[ x_j - \langle x_j \rangle \right] \rangle}
{\sqrt{\langle \left[ x_i - \langle x_i \rangle \right] ^2 \rangle \langle \left[ x_j -\langle x_j \rangle \right] ^2 \rangle}}
\end{equation}
where $x_{i}$ and $x_{j}$ represent histogram counts in bins $i$ and $j$ and $\langle x_{i}\rangle$ represents the average value of variable $x_{i}$ over all traces. The numerator in Eq.~\ref{eq:2DCH} calculates the average covariance between conductance bins $i$ and $j$. The denominator scales the numerator so that the values of $C$ range from $[-1,1]$. When $C_{i,j}=0$ there is no correlation between conductance bins $i$ and $j$. When $C_{i,j}>0$ ($C_{i,j}<0$) there is a correlation (anticorrelation) between conductances corresponding to bins $i$ and $j$. Generally, conductance bins $i$ and $j$ are correlated when there is a probability of plateaus occurring in conductance traces at both conductances together in the same trace. Anticorrelation may indicate that there is a probability that if a plateau occurs at conductance bin $i$, then there will be no plateau at conductance bin $j$. The next step in this method requires, after calculating $C$, to make certain assumptions about the trajectories of the break junctions, and then separate the data into groups based on those assumptions. Our new approach does not require these assumptions.

The established method of principal component analysis (PCA)\cite{Pearson1901, Jolliffe2002, Shlens2014,Bro2014} provides a statistically rigorous, objective tool for sorting data sets. PCA is a common method applied to a variety of disciplines including neural networks\cite{Karhunen1994}, chemometrics\cite{Bro2014}, and geospatial statistics\cite{Bevington2016}.

A detailed summary of the relevant mathematics behind PCA can be found in the Supporting Materials (SM). In short, diagonalizing $C$, Eq.~\ref{eq:2DCH}, and sorting the eigenvectors in decreasing size of eigenvalues yields an orthogonal basis set with the first eigenvector, $PC_1$, describing the direction of most variance in the data set, and the second eigenvector, $PC_2$, describing the second most variance, etc.

\begin{figure}
\includegraphics[width=8.6cm]{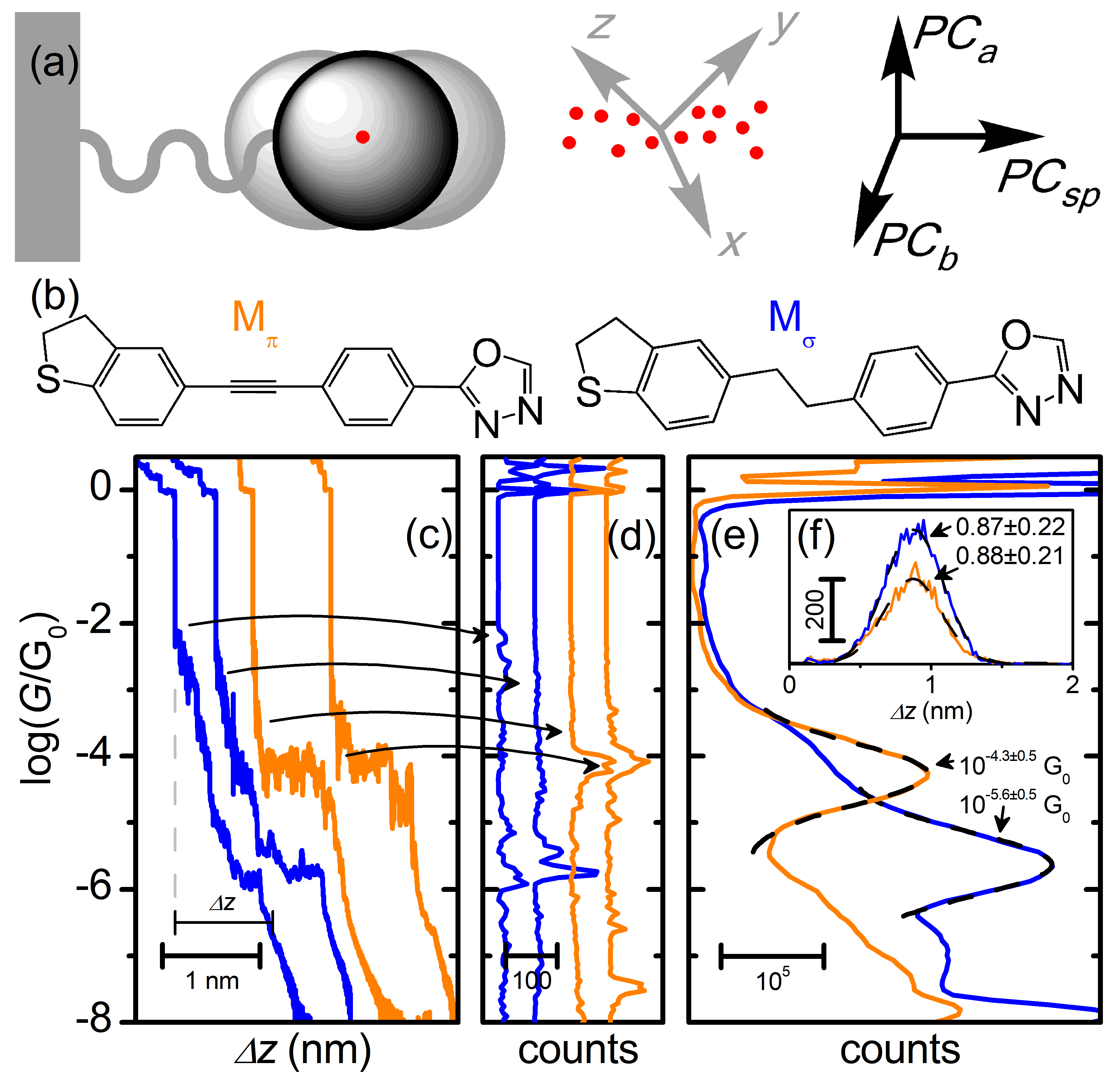}
\caption{\label{fig:one} (a) PCA demonstration schematic; and break junction results for
(b) $\mathbf{M}_\pi$ (orange) and $\mathbf{M}_\sigma$ (blue),
(c) example conductance traces,
(d) example conductance traces binned into single 1D histograms,
(e) total 1D histograms accumulated from $86\%$ of $8801$ ($\mathbf{M}_\pi$) and $96\%$ of $12033$ traces ($\mathbf{M}_\sigma$) - molecular peaks fit to Gaussian ( curves),
(f) plateau length histograms calculated as the displacement of each trace between $10^{-0.3}~\textrm{G}_0$ and $10^{-6.5}~\textrm{G}_0$.}
\end{figure}

An intuitive understanding of PCA can be obtained by reducing the problem to a ball on a spring oscillating in three dimensions [Fig.~\ref{fig:one}(a)]. Although the ball is oscillating in a particular direction due to the large influence of the spring, there will be smaller influences in other directions. $C$ will be a $3\times3$ matrix describing how the three variables correlate with one another. Diagonalizing $C$ will result in three eigenvalue-eigenvector pairs. The eigenvector corresponding to the largest eigenvalue, $PC_{sp}$, will point along the direction of the oscillating spring. The other eigenvectors, $PC_a$ and $PC_b$, will point orthogonal to the spring oscillations, in directions corresponding to smaller influences.

This basic understanding applies to the following analysis, although it will be carried out in 128 dimensions. To sort traces, the procedure will be to project all measurements onto one PC and sort them depending on whether they project positively or negatively. In the above analogy, this is equivalent to sorting the measurements into groups based on whether they are on the elongated part of the spring motion or the compressed part.

To test the ability for PCs to effectively sort SMBJ data sets into single mode distributions, two molecules, $\mathbf{M}_\pi$ and $\mathbf{M}_\sigma$ [Fig.~\ref{fig:one}(b)], were synthesized with identical anchoring groups, similar lengths, but different conductances due to the $\pi$ or $\sigma$ bridge (see SM for synthesis details). The molecules were chosen so that a mixture of the molecules can be measured using SMBJs (see SM for SMBJ experimental details) and both will anchor identically and compete equally for the break junction. The molecules were also chosen because there was no expected specific interaction between $\mathbf{M}_\pi$ and $\mathbf{M}_\sigma$ when mixed in solution.

First $\mathbf{M}_\pi$ and $\mathbf{M}_\sigma$ were measured separately with SMBJ. Traces [Fig.~\ref{fig:one}(c)] showed a difference in conductance in the molecular plateau between $\mathbf{M}_\pi$ and $\mathbf{M}_\sigma$. The conductance traces were then binned into 1D histograms [Fig.~\ref{fig:one}(d)]. The data matrices, $\mathbf{X}_\pi$ and $\mathbf{X}_\sigma$, had single 1D histograms as columns, and histogram bins as rows. The sum of each row produced total 1D histograms [Fig.~\ref{fig:one}(e)]. $\mathbf{X}_\pi$ and $\mathbf{X}_\sigma$ were easily distinguishable by the change in location of the molecular conductance peak from $10^{-4.3}~\textrm{G}_0$ for $\mathbf{M}_\pi$, to $10^{-5.6}~\textrm{G}_0$ for $\mathbf{M}_\sigma$, as determined by fitting the peaks to a Gaussian. Although the molecular conductances changed by over one order of magnitude, the plateau lengths, accumulated into plateau length histograms [Fig.~\ref{fig:one}(f)], showed no difference in the average plateau length between $\mathbf{M}_\pi$ and $\mathbf{M}_\sigma$.

\begin{figure}
\includegraphics[width=8.6cm]{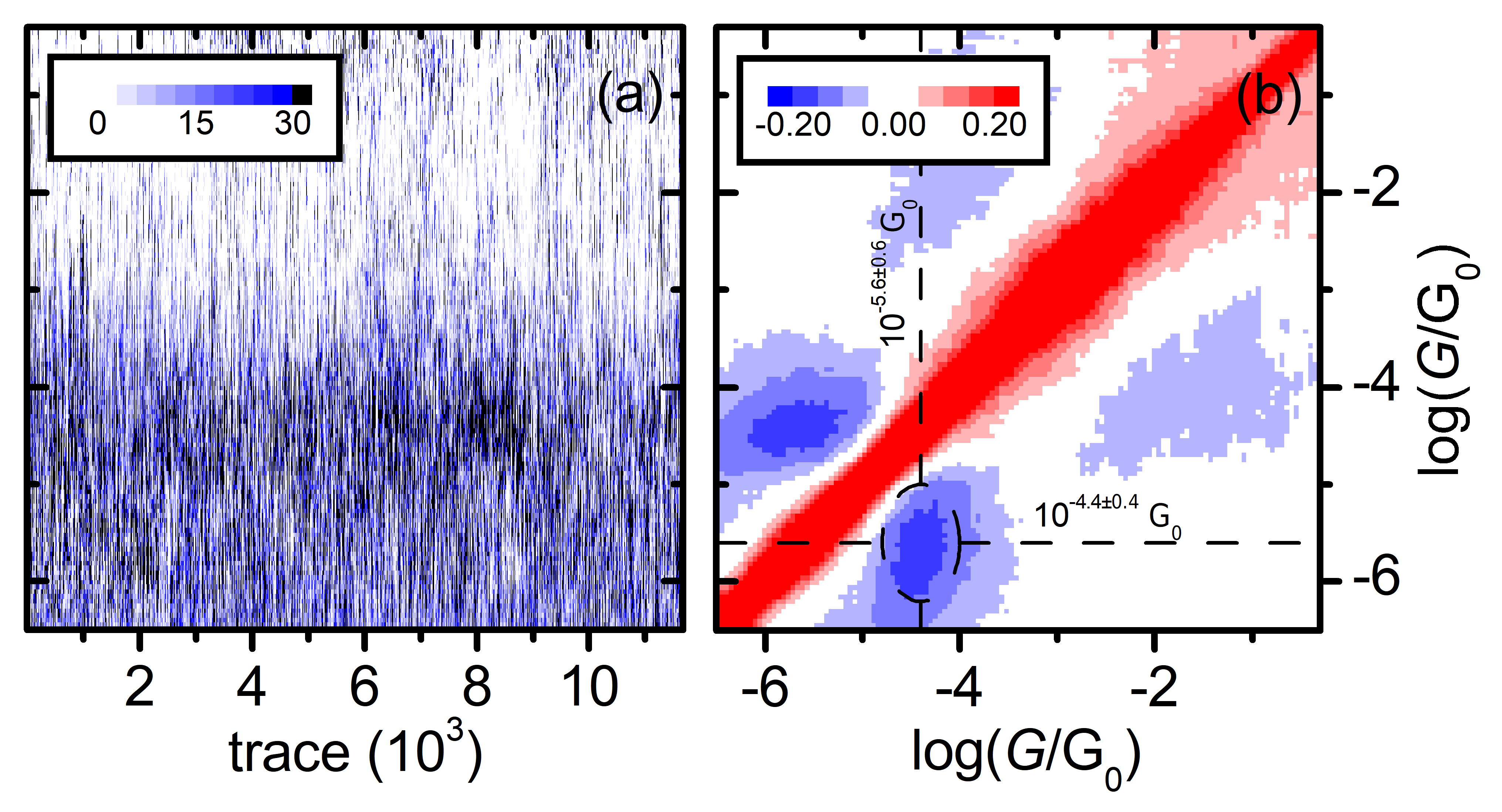}
\caption{\label{fig:two} Break junction results for \textbf{Mix 1}.
(a) Data matrix of $n=11671$ traces binned into single histograms of $m=128$ bins;
(b) 2D correlation histogram calculated from data matrix.}
\end{figure}

Next a 1:3 (\#/\#) mixture of $\mathbf{M}_\pi$ and $\mathbf{M}_\sigma$ (\textbf{Mix 1}) was measured (see SM for results for two other mixtures). Due to the differences in conductance between $\mathbf{M}_\pi$ and $\mathbf{M}_\sigma$, we anticipated a bimodal distribution in this first example. To test our method, we needed to show PC sorting can distinguish the two classes of events which are contributing to this bimodal distribution. 1D histograms were created from each trace, and these were accumulated into a data matrix [Fig.~\ref{fig:two}(a)]. The bins corresponding to the gold-gold junction and open circuit were removed and the correlation matrix was calculated. An intensity plot of the correlation matrix was plotted [Fig.~\ref{fig:two}(b)]. Figure~\ref{fig:two}(b) had a region of anticorrelation at $(10^{-4.3}~\textrm{G}_0,10^{-5.6}~\textrm{G}_0)$ which suggested that the traces in the data set had a plateau at the molecular conductance of $\mathbf{M}_\pi$ or $\mathbf{M}_\sigma$, but not both.

Following the guidelines outlined in Ref.~\cite{Makk2012}, the next steps in analyzing \textbf{Mix 1} would ordinarily require formulating a hypothesis about the expected shape of the conductance traces that yields the anticorrelation region, and then sorting the traces into groups using these conditions. The method proposed in this Letter forgoes these assumptions.

Instead the correlation matrix was diagonalized and the eigenvalues were sorted in decreasing order. The corresponding eigenvectors were sorted with the eigenvalues. 

\begin{figure}
\includegraphics[width=8.6cm]{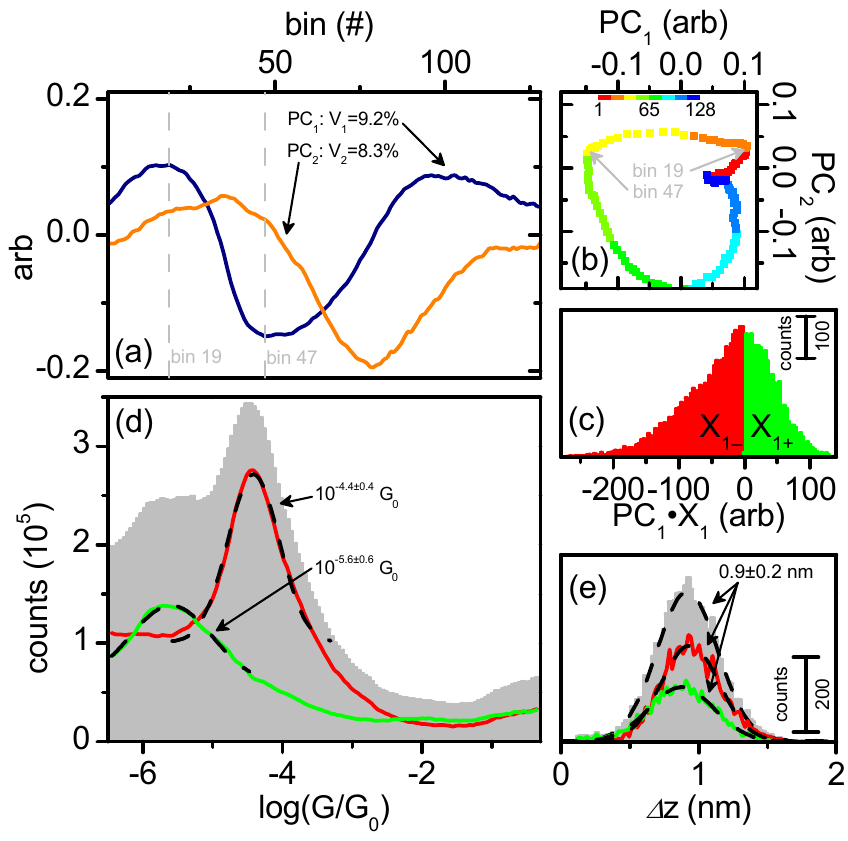}
\caption{\label{fig:three} Principal component sorting on \textbf{Mix 1} using $PC_1$.
(a) $PC_1$ (blue) and $PC_2$ (orange) plotted vs. bin number;
(b) $PC_1$ vs. $PC_2$ from bin 1 (red) to bin 128 (blue);
(c) histogram constructed from single 1D histograms projected onto $PC_1$ - curves in the red (green) bins had negative (positive) projections;
(d) negatively (7129 curves) and positively (4541 curves) projected curves sorted into separate groups and total (80\% of 14543 curves) 1D histograms (gray);
(e) plateau lengths were determined for each trace in the negative (red) and positive (green) subgroups and the entire data set (gray) and histograms were constructed for each.}
\end{figure}

The question of which eigenvectors to use in further analysis, or how many, is a poorly resolved issue.\cite{Bro2014} A variety of criteria are suggested with the caveat that, regardless of method, one must test and check based on the results with which to proceed. For our study, it was necessary to find the minimum number of PCs with the highest variance explained that distinguished between two different conductance features. We found that either the first or second eigenvector, named $PC_1$ and $PC_2$, respectively [Fig.~\ref{fig:three}(a)], was sufficient to do this. More PCs may contain other details about the data set, but these details were not the focus of this study. To determine which, $PC_1$ or $PC_2$, to use in the final steps of the sorting, two criteria were used: \begin{enumerate*}\item a score plot, $PC_1$ vs. $PC_2$ [Fig.~\ref{fig:three}(b)] aided in understanding the relative importance of $PC_1$ and $PC_2$ with respect to each other; and \item the PC had to work\end{enumerate*}. Criteria 1 aided in objectively choosing the appropriate PC because we were predominantly concerned with a PC which found the largest distinctions between single 1D histograms with features either associated with $\mathbf{M}_\pi$ or $\mathbf{M}_\sigma$. When the bin coefficients in Fig.~\ref{fig:three}(b) for either $PC_1$ or $PC_2$ are close to zero then those bins do not contribute to that PC. On the other hand, when the bin coefficients for a PC are large, either positively or negatively, then those bins are significant for that PC. Thus, large coefficients in the region of the molecular conductance and small coefficients elsewhere in the PC is desirable. But since the toolbox of PCA still does not provide a definitive test to choose PCs, criteria 2 was also necessary.  Criteria 2 had the potential to be a subjective criteria, but it was very clear when the PC successfully sorted the data, and when it did not, and in every case either $PC_1$ or $PC_2$ did this effectively. When the PC successfully sorted the traces, the total 1D histograms for each group were single mode Gaussian distributions with little or no shoulders. When the PC did not successfully sort the traces, the 1D histograms retained a bimodal distribution or large shoulders. If neither $PC_1$ nor $PC_2$ fulfilled either criteria 1 or criteria 2, it would be necessary to move on to other PCs. PCs after $PC_2$ (see Fig. S9 in SM for the first eight PCs) followed a qualitatively different pattern than PCs $1$ and $2$. The variances explained declined rapidly, suggesting the first PCs were more significant than the others (see SM for discussion).

For the example of \textbf{Mix 1}, bins $19$ and $47$ were the bins where $PC_1$ was maximum and minimum, and closely corresponded to bins with conductances of $\mathbf{M}_\sigma$ and $\mathbf{M}_\pi$, respectively, so $PC_1$ was chosen to complete the analysis. In this case $PC_2$ will not work because it was maximum at $10^{-2}~\textrm{G}_0$ and this is in a region outside the range of the molecular conductances. This region shows up in Fig.~\ref{fig:two}(b) as the weak anticorrelation region at $(10^{-2}~\textrm{G}_0,10^{-4}~\textrm{G}_0)$ (see SM for PC sorting using $PC_2$). $PC_1$ accounted for over 9\% of the variance of the entire data set, while the remaining 127 eigenvectors together accounted for the remaining 91\%. Studying $PC_1$ alone retains much of the important variability of the data, while significantly reducing the complexity of the analysis by focusing on one dimension. If more PCs were used to create a multidimensional feature vector it was possible to over-fit the results. For the current Letter we chose the smallest number of PCs that still yielded a reasonable result, and a single PC was sufficient to do this. It is remarkable that 9\% variance explained is indeed sufficient to distinguish between different conductance features in our data sets. For other systems, for instance systems with larger signal to noise ratios, a larger variance explained may be appropriate.

Next, each 1D histogram was projected onto $PC_1$ and the 1D histograms were separated into positive and negative groups, $\mathbf{X}_{1+}$ (4541 curves) and  $\mathbf{X}_{1-}$ (7129 curves) respectively, based on the sign ($+/-$) of the dot product [Fig.~\ref{fig:three}(c)]. Each group was separately summed into total 1D histograms [Fig.~\ref{fig:three}(d)] and compared to Fig.~\ref{fig:one}(d). The histograms were separated into high conductance ($\mathbf{X}_{1-}$) and low conductance groups ($\mathbf{X}_{1+}$), corresponding to junctions involving $\mathbf{M}_\pi$ and $\mathbf{M}_\sigma$, respectively, because the group histograms retained none of the bimodal nature that the full data set's histogram did. Finally, the length histograms for the sorted groups were compared to the length histograms for the entire set [Fig.~\ref{fig:three}(e)]. The average plateau lengths for each group were the same, and matched both the average plateau length of the entire set, and the average plateau lengths when the molecules were measured separately [Fig.~\ref{fig:one}(e)].

As proof of concept, the results above showed the ability of PC sorting to distinguish obvious bimodal features in data sets with a change in conductance of over one order magnitude between $\mathbf{M}_\pi$ and $\mathbf{M}_\sigma$. Next we show that PC sorting can perform a very useful task: distinguish groups with a change in conductance of about a half order of magnitude and an associated change in plateau length of about $0.15~\textrm{nm}$. Changes in conductance like this are common in optically\cite{Tebikachew2017} and electrochemically\cite{Aragones2017Metal-Controlled} switched SMBJs. Furthermore, $0.15~\textrm{nm}$ represents the radius of a gold atom and the bottom limit of measurable differences between groups.

\begin{figure}
\includegraphics[width=8.6cm]{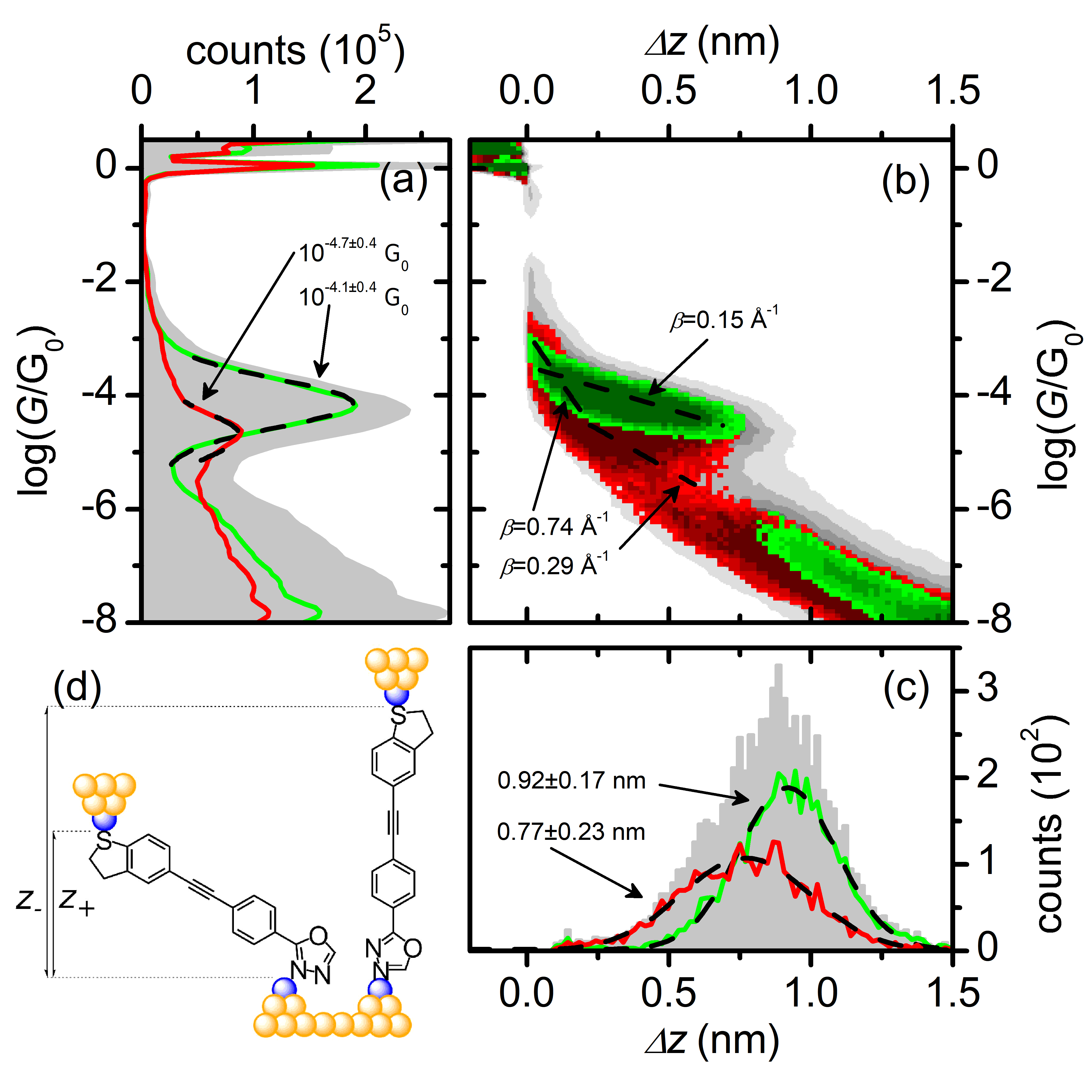}
\caption{\label{fig:four} PC sorting of $\mathbf{M}_\pi$ data set (86\% of 8801 curves) into positive ($\mathbf{X}_{\pi +}$, green, 3151 curves) and negative ($\mathbf{X}_{\pi -}$, red, 4419 curves) subgroups using $PC_2$.
(a) total 1D histograms with Gaussian fits;
(b) 2D histograms with master curve linear fit of plateaus;
(c) plateau length histograms with Gaussian fits;
(d) cartoon depicting possible junction geometry leading to $\mathbf{X}_{\pi -}$ and $\mathbf{X}_{\pi +}$.}
\end{figure}

The SMBJ results for $\mathbf{M}_\pi$ were analyzed using the same procedures outlined above. $PC_2$ was used to separate the single histograms into positive ($\mathbf{X}_{\pi -}$) and negative ($\mathbf{X}_{\pi +}$) groups (Fig.~\ref{fig:four}, see SM for PC sorting steps). $\mathbf{X}_{\pi +}$ had a larger molecular conductance in the 1D histogram [Fig.~\ref{fig:four}(a)] and a longer plateau length [Fig.~\ref{fig:four}(c)] compared to $\mathbf{X}_{\pi -}$. Most traces in $\mathbf{X}_{\pi +}$ had long, flat plateaus resulting in a narrow total 1D histogram. The average master curves were calculated to visualize an average trace  (see SM for details). The slope, $\beta$, of the plateau region in the master curve for $\mathbf{X}_{\pi +}$ was determined to be $0.15~\textrm{\AA}^{-1}$ with a linear fit. Traces in $\mathbf{X}_{\pi -}$ were less homogeneous, with fewer long and flat plateaus and a broader distribution of steeper slopes. The slope of the master curve in the plateau region of group $\mathbf{X}_{\pi -}$ was also drastically different: the master curve had two regions in the molecular plateau with two separate slopes, $0.74$ and $0.29~\textrm{\AA}^{-1}$. The slope of the plateau of a SMBJ trace was shown to be proportional to the tunneling decay constant of the Au-molecule-Au system.\cite{Huang2015break} The gold-molecule-gold system can be modeled as a square potential barrier yielding an exponential dependence of tunneling conductance with length:
\begin{equation}
G=A\exp{(-\beta l)}
\end{equation}
where $A$ captured parameters in the contacts and $l$ was the width of the potential well.\cite{Simmons1963generalized} Thus, the doubling of the tunneling decay constant between $\mathbf{X}_{\pi +}$ and $\mathbf{X}_{\pi -}$ reflected an important change in the tunneling behavior of $\mathbf{M}_\pi$. An average slope of $0.15~\textrm{\AA}^{-1}$ for $\mathbf{X}_{\pi +}$ matched other conjugated molecules which were shown to have slopes of $0.1-0.4\textrm{\AA}^{-1}$. Likewise, an average slope of $0.74~\textrm{\AA}^{-1}$ for $\mathbf{X}_{\pi -}$ matched saturated molecules which were shown to have slopes of $0.6-1.0\textrm{\AA}^{-1}$.\cite{Huang2015break}

Because the sorting process was objective and statistically relevant, we were encouraged to speculate about the physical differences leading to the sorted groups. If the sorting was a result of decisions about expected trace shapes, assertions about the physical interpretations of the data will likely reinforce the subjective decisions applied. Instead, PC sorting allows the experimentalist to make confident hypotheses which can better inform new directions of investigation, including simulations to test the hypotheses.

With this in mind, we attempt here to interpret the results of the PC sorting on $\mathbf{M_\pi}$. $PC_2$ had the second largest variance explained of the 128 eigenvectors of $\mathbf{X_\pi}$, accounting for 8\% of the variance. We hypothesized $\mathbf{X}_{\pi +}$ comprised mostly of junctions in which the end groups of $\mathbf{M}_\pi$ bonded strongly and bridged the electrodes in a nearly perpendicular geometry. This geometry yielded a high conductance, low tunneling decay constant, and long plateau length. We further hypothesized $\mathbf{X}_{\pi -}$ comprised of junctions in which the 1,3,4-oxadiazole end group of $\mathbf{M}_\pi$ was more weakly bonded to one electrode, most likely slipped along the electrode (thus achieving at least two metastable geometries responsible for two distinct tunneling decay regions), and never achieved a full chemisorbed bond with the electrode, which would have resulted in identical conductance and length as $\mathbf{X}_{\pi+}$. It was shown\cite{Kamenetska2009} that a bonding geometry in which the conductance pathway does not align with the Au electrode will yield a lower conductance.

The PC sorting method described in this Letter provided a means to sort a large data set into groups based on statistically distinctive characteristics, allowing the groups to be studied separately and allowed two particular groups to be compared meaningfully to theoretical predictions based on models of ideal junctions. No \textit{a priori} hypothesis needed to be imposed in regards to expected conductance trace shape. In this Letter the PCs were treated separately, yielding an intrinsically 1D analysis by comparing either/or behavior of the data set. Presented here was a technique which can be applied to many data sets across many disciplines. Many variables can be included in a data matrix, including measurements from conducting force spectroscopy and optical measurements. Also, multiple PCs may be treated as a single feature vector - the result would be a PC sorting based on higher dimensional, more complex criteria.

\begin{acknowledgments}
The authors would like to dedicate this work to the memory of Dr. Thomas Wandlowski. The authors are grateful to Dr. Wenjing Hong for design of molecules. J.M.H is grateful to James Bevington and Veerabhadrarao Kaliginedi for illuminating discussions. This work was generously supported by the EC FP7 ITN ``MOLESCO'' project number 606728, and the University of Bern.
\end{acknowledgments}


\bibliographystyle{apsrev4-1}

%

\end{document}


\title{Supporting Material: \\ Fast data sorting with modified principal component analysis to distinguish unique single molecular break junction trajectories}%

\author{J.M.~Hamill} \affiliation{Department of Chemistry and Biochemistry, University of Bern, Freiestrasse 3, CH-3012, Bern, Switzerland}
\author{X.T.~Zhao} \affiliation{Department of Chemistry, Durham University, Durham DH1 3LE, United Kingdom}
\author{G.~M\'{e}sz\'{a}ros} \affiliation{Research Centre for Natural Sciences, Hungarian Academy of Sciences, Magyar Tud\'{o}sok K\"{o}r\'{u}tja 2, H-1117 Budapest, Hungary}
\author{M.R.~Bryce} \affiliation{Department of Chemistry, Durham University, Durham DH1 3LE, United Kingdom}
\author{M.~Arenz} \affiliation{Department of Chemistry and Biochemistry, University of Bern, Freiestrasse 3, CH-3012, Bern, Switzerland}
\email[Author email: ]{joseph.hamill@dcb.unibe.ch}
\date{\today}%
\maketitle
\tableofcontents

\section{Principal components as description of variance}
The relationships between the eigenvalues, eigenvectors, and variance in a correlation matrix were demonstrated by Pearson\cite{Pearson1901}. Here we wish to summarize the relevant mathematics and notation used in the main text.

Given an $m\times n$ data matrix, $\mathbf{X}$, with $m$ variables and $n$ measurements, we wish to find a transformation, $\mathbf{a}_k$, such that 
\begin{equation}\label{eq:p}
\mathbf{p}_k=\mathbf{a}_k^T \mathbf{X},~k=1,\hdots,m.
\end{equation}
Specifically, we wish to choose $\mathbf{a}_k$ such that 
\begin{equation}\label{eq:var_p}
var \left[ \mathbf{p}_k \right]=\langle \mathbf{p}^2_k \rangle - \langle \mathbf{p}_k \rangle ^2
\end{equation}
 is maximized, and
\begin{equation}\label{eq:cov_p}
cov \left[\mathbf{p}_k, \mathbf{p}_l \right] = 0,~k,l=1,\hdots,m,~k>l\geq1.
\end{equation}
To determine the form of $\mathbf{a}_k$, Eq.~\ref{eq:var_p} can be expanded

\begin{align*}
var \left[ \mathbf{p}_k \right] &= \langle \mathbf{p}^2_k \rangle - \langle \mathbf{p}_k \rangle ^2 \\
 &=\sum_{i,j=1}^{n} a_{ik}a_{jk} \langle x_i x_j \rangle - \sum_{i,j=1}^{n} a_{ik}a_{jk} \langle x_i \rangle \langle x_j \rangle \\ 
 &= \sum_{i,j=1}^{n} a_{ik}a_{jk}S_{ij},~S_{ij} \equiv \sigma_{x_i x_j} = \langle x_i x_j \rangle - \langle x_i \rangle \langle x_j \rangle\\ 
 &= \mathbf{a}_k^T \mathbf{S} \mathbf{a}.
\end{align*}

$\mathbf{S}$ is the covariance matrix for the data matrix $\mathbf{X}$. To find $\mathbf{a}_k$ which maximizes $var[\mathbf{p}_k]$, subject to $\mathbf{a}_k^T\mathbf{a}_k = 1$, we will use the method of Lagrangian multipliers. This amounts to maximizing
\begin{equation*}
\mathbf{a}_k^T\mathbf{S} \mathbf{a}_k - \lambda (\mathbf{a}_k^T\mathbf{a}_k-1)
\end{equation*}
by differentiating and finding the zeros
\begin{align*}
\mathbf{S}\mathbf{a}_k - \lambda \mathbf{a}_k &= 0\\
(\mathbf{S} - \lambda \mathbf{I}_m) \mathbf{a}_k &= 0.
\end{align*}
Evidently $\mathbf{a}_k$ is an eigenvector of the covariance matrix, $\mathbf{S}$, and $\lambda_k$ is the corresponding eigenvalue. Therefore we also know
\begin{align*}
var \left[ \mathbf{p}_1 \right] &= \mathbf{a}_1^T \mathbf{S} \mathbf{a}_1 \\
&= \mathbf{a}_1^T \lambda_1 \mathbf{a}_1 \\
&=\lambda_1.
\end{align*}
Therefore $\lambda_1$ is the largest eigenvalue and $\mathbf{a}_1$, the first principal component (PC), projects the data matrix $\mathbf{X}$ along the direction which maximizes the variance. Furthermore, $\lambda_2$ is the second largest eigenvalue (Eq.~\ref{eq:cov_p}) and $\mathbf{a}_2$, the second PC, projects the data matrix $\mathbf{X}$ in a direction, orthogonal to $\mathbf{a}_1$, which achieves the second largest variance. Specifically, $sign \left[ p_{1i} \right]$ describes whether measurement $i$ more closely resembles $\mathbf{a}_{1}$ or $-\mathbf{a}_{1}$. This will be used as the sorting criteria to separate the following data sets into groups. 

Finally, all of the eigenvectors together account for the entirety of the variance within the data set, and
\begin{equation*}
V_k= \frac{\lambda_k}{\sum_{k=1}^{m} \lambda_k} \cdot 100\%
\end{equation*}
meaning each eigenvalue describes the amount, given as a percentage of the total, of variance explained, $V_k$. If the data matrix comprised entirely of randomly distributed data, then the covariance matrix would be unity along the diagonal, and zero elsewhere. In other words, no other variables would be correlated at all, except with themselves. In this case, the eigenvalues, and variance explained, would be degenerate, with value of unity. This is the one extreme, when the data is random. The other extreme is when the data is all perfectly correlated. In this case, one PC would be sufficient to describe 100\% of the variance. Thus, the degree to which the variance explained falls off is a good description of the amount of randomness is in the data set. If the variance explained versus PC number is nearly flat, it will indicate there is a lot of randomness in the data, and if it is steep it will indicate that there is a strong trend in the data.

Normally it is necessary to normalize each variable and offset the mean to zero, thus giving each variable equal scaling, before calculating $\mathbf{S}$. However, for our purposes it is more desirable to calculate the correlation matrix, $C$, introduced by Halbritter, \textit{et al.}\cite{Halbritter2010, Makk2012} than the covariance matrix, $\mathbf{S}$
\begin{equation}\label{eq:2DCH}
C_{i,j}=\frac
{\langle \left[ x_i - \langle x_i \rangle \right] * \left[ x_j - \langle x_j \rangle \right] \rangle}
{\sqrt{\langle \left[ x_i - \langle x_i \rangle \right] ^2 \rangle \langle \left[ x_j -\langle x_j \rangle \right] ^2 \rangle}}.
\end{equation}
because the denominator ensures each variable has been offset and normalized. In general, covariance between different variables in a data matrix can be misleading, because the different variables may have different units at vastly different scales. The denominator in Eq.~\ref{eq:2DCH} removes this vagueness by removing the units and normalizing all variables to vary in the range from $[-1,1]$. It is therefore useful to use Eq.~\ref{eq:2DCH} for two reasons: \begin{enumerate*}\item analysis using the correlation matrix has become a standard analysis method in break junction research, and \item the normalized nature of the correlation matrix allows it to be immediately utilized in principal componant analysis methods.\end{enumerate*}
\section{Mechanically controlled break junctions}
A mechanically controlled break junction (MCBJ) was used to create single molecular break junctions. MCBJ utilizes a piezo stack to rapidly open and close a $0.1~\textrm{mm}$ diameter Au wire (Goodfellow, 99.99\%) inside a liquid cell filled with the molecular solution. Each trace [example traces in Fig. 1(c)] started at high conductance with a closed Au-Au circuit. After the last Au-Au atomic contact broke, the conductance dropped rapidly from $10^0~\textrm{G}_0$. An ideal junction which was mechanically stable and free of contamination had a period of snap-back just following the breaking of the Au-Au bond in which the conductance dropped so quickly that the trace showed no significant change in displacement while the conductance dropped by two or three orders of magnitude.\cite{Huang2015break} If a molecule was trapped in the junction, a plateau was seen in the trace while the conductance across the molecule was measured. Finally, the Au-molecule-Au junction broke and the conductance dropped to the open circuit sensitivity of the amplifier, at approximately $10^{-9}~\textrm{G}_0$. The separation of the Au leads, $\Delta z$, was deduced from the length of the molecular plateau feature, from after the Au-Au junction to before the open circuit.

A very permissive pre-analysis selection criteria was applied to remove traces which did not meet the following criteria:
\begin{enumerate}
\item start above a conductance of at least $10^{0.5}~\textrm{G}_0$ - this ensured that each junction began with a Au-Au contact
\item end below the open circuit sensitivity of the amplifier, at $10^{-9}~\textrm{G}_0$ - this ensured that each trace opened completely
\item the trace had to be shorter than a maximum length of $6~\textrm{nm}$ - traces longer than this do not involve the molecule of study ($1.6~\textrm{nm}$ long), but may be due to contamination or environmental disturbances.
\end{enumerate}
This criteria removed between $2$ and $20\%$ of the traces in a data set, depending on how stable the experiment was.

\section{Synthetic procedures}
All commercial chemicals were used without further purification. Anhydrous solvents were dried through an HPLC column on an Innovative Technology Inc. solvent purification system. Column chromatography was carried out using 40-60 $\mu$m mesh silica (Fluorochem). NMR spectra were recorded on: Bruker Avance-400, Varian VNMRS-600, VNMRS-700 and Varian Inova 500 spectrometers. Chemical shifts are reported in ppm relative to tetramethylsilane (0.00 ppm). Melting points were determined in open-ended capillaries using a Stuart Scientific SMP40 melting point apparatus at a ramping rate of 2 \si{\degreeCelsius}/min. Mass spectra were measured on a Waters Xevo OTofMS with an ASAP probe. Electron ionization (EI) mass spectra were recorded on a Thermoquest Trace or a Thermo-Finnigan DSQ.
\begin{figure}
\includegraphics[width=8.6cm]{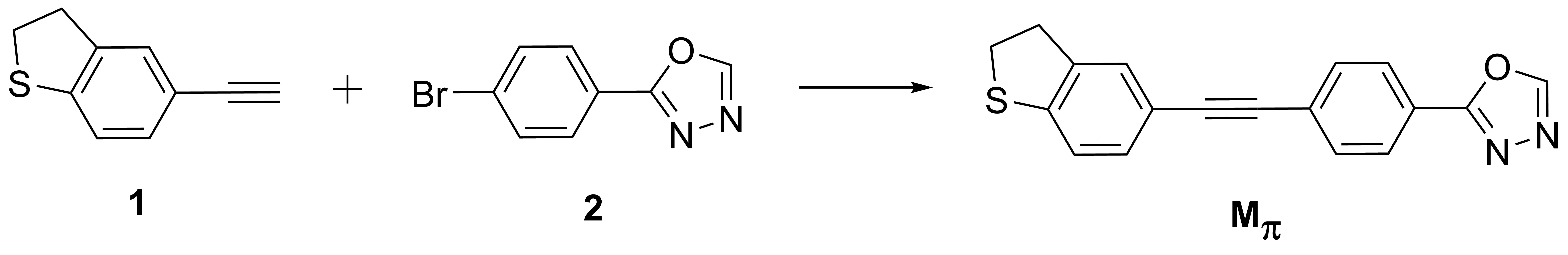}
\caption{\label{fig:s_mpi} Synthesis of $\mathbf{M}_\pi$. Reagents and conditions: \ce{Pd(PPh3)4}, \ce{CuI}, THF/\ce{Et3N}, 18 h, 60 \si{\degreeCelsius}, 64\%.}
\end{figure}

To a solution of \textbf{1}\cite{Moreno-Garcia2013} (Fig.~\ref{fig:s_mpi}, 100 mg, 0.63 mmol), \ce{Pd(PPh3)4}  (36 mg, 0.03 mmol) and \ce{CuI} (6 mg, 0.03 mmol) in \ce{THF/Et3N} (3:1 v/v) (25 ml), was added \textbf{2}\cite{Salvanna2013} (Fig.~\ref{fig:s_mpi}, 142 mg, 0.63 mmol) and the reaction was stirred at 60 \si{\degreeCelsius} for 18 h under argon (which had been dried by passage through a column of phosphorus pentoxide)  before the solvent was removed by vacuum evaporation and the residue was purified by column chromatography in \ce{DCM} which gave $\mathbf{M}_\pi$ as a yellow solid (122 mg, 64\% yield). m.p.: decomp. over 250 \si{\degreeCelsius}. HR-MS (ASAP+) \textit{m/z} Calcd for \ce{C18H13N2OS} [M+H]$^+$ 305.0749, found \textit{m/z} : [M+H]$^+$ 305.0753. Anal. Calc. for \ce{C18H12N2O}: C, 71.03; H, 3.97; N, 9.20. Found: C, 71.13; H, 3.92; N, 9.11. \ce{^1H} NMR (Fig.~\ref{fig:s_NMR1}, 400 MHz, \ce{CDCl3}) $\delta$ 8.48 (s, 1H), 8.06 (d,  \textit{J} = 8.8, 2H), 7.64 (d, \textit{J} = 8.8, 2H), 7.36 (m, 1H), 7.31 (m, 1H), 7.20 (m, 1H), 3.40 (t, \textit{J} = 7.8 Hz, 2H),  3.30 (t, \textit{J} = 7.8 Hz, 2H). \ce{^13C} NMR (Fig.~\ref{fig:s_NMR2}, 101 MHz, \ce{CDCl3}) $\delta$ 164.61, 152.88, 143.73, 140.71, 132.27, 131.19, 127.60, 127.54, 127.21, 122.74, 122.25, 118.30, 93.12, 88.22, 36.04, 33.69.
\begin{figure}
\includegraphics[width=8.6cm]{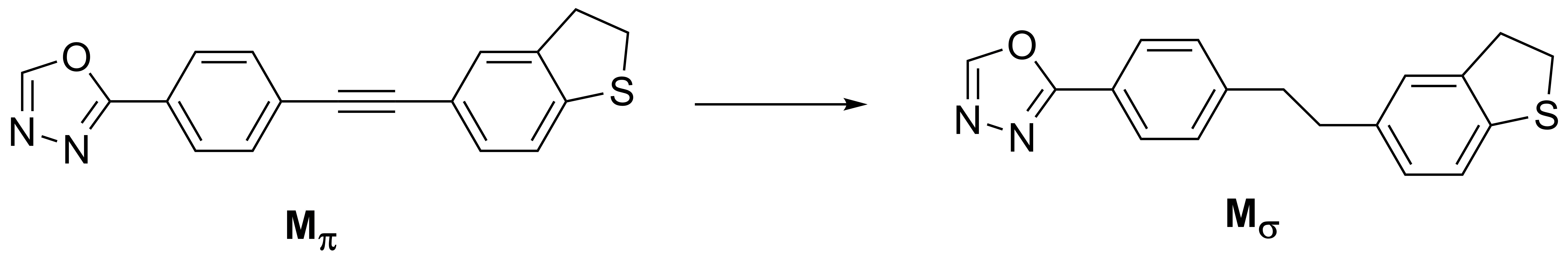}
\caption{\label{fig:s_msi} Synthesis of $\mathbf{M}_\sigma$. Reagents and conditions: Pd/C, THF, 24 h, r.t., 65\%.}
\end{figure}

A mixture of $\mathbf{M}_\pi$  (25 mg, 0.08 mg) and \ce{Pd/C} (10 mg, 10wt. \%) in \ce{THF} under \ce{H2} was stirred at r.t. for 24 h before the mixture was filtered through a silica pad. A yellow product $\mathbf{M}_\sigma$ (Fig.~\ref{fig:s_msi}, 16 mg, 65\%) was obtained from recrystallisation in \ce{DCM/EtOH}. m.p.: decomp. over 265 \si{\degreeCelsius}. HR-MS (ASAP+) \textit{m/z} Calcd for \ce{C18H17N2OS} [M+H]$^+$ 309.1062, found \textit{m/z} : [M+H]$^+$ 309.1073. Anal. Calc. for \ce{C18H16N2OS}: C, 70.10; H, 5.23; N, 9.08. Found: C, 70.24; H,5.30; N, 9.20. \ce{^1H} NMR (Fig.~\ref{fig:s_NMR3}, 400 MHz, \ce{CDCl3}) $\delta$ 8.47 (s, 1H), 8.01 (d, \textit{J} = 8.3 Hz, 2H), 7.33 (d, \textit{J} = 8.3 Hz, 2H), 7.14 (d, \textit{J} = 7.8 Hz, 1H), 7.02 (s, 1H), 6.93 (d, \textit{J} = 7.8, 1H), 3.38 (t, \textit{J} = 7.6, 1.3 Hz, 2H), 3.26 (t, \textit{J} = 7.6, 1.3 Hz, 2H), 2.99 (m, 2H), 2.90 (m, 2H). \ce{^13C} NMR (Fig.~\ref{fig:s_NMR4}, 101 MHz, \ce{CDCl3}) $\delta$ 152.64, 146.42, 140.59, 139.19, 137.33, 129.53, 127.74, 127.37, 124.89, 122.19, 121.44, 38.31, 37.24, 36.38, 33.71.
\begin{figure}
\includegraphics[width=8.6cm]{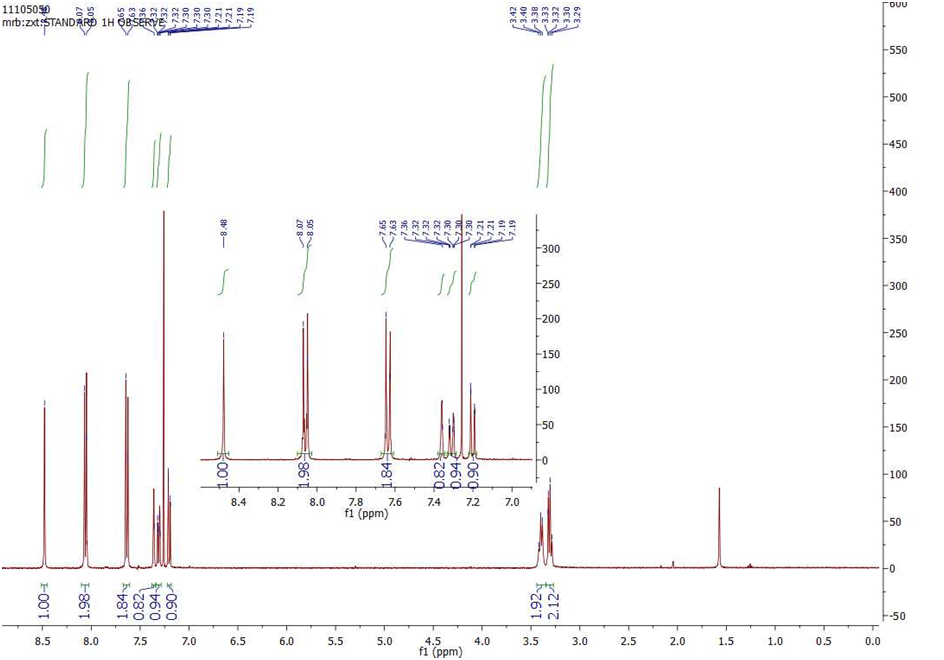}
\caption{\label{fig:s_NMR1}\ce{H^1} NMR Spectrum for $\mathbf{M}_\pi$.}
\end{figure}
\begin{figure}
\includegraphics[width=8.6cm]{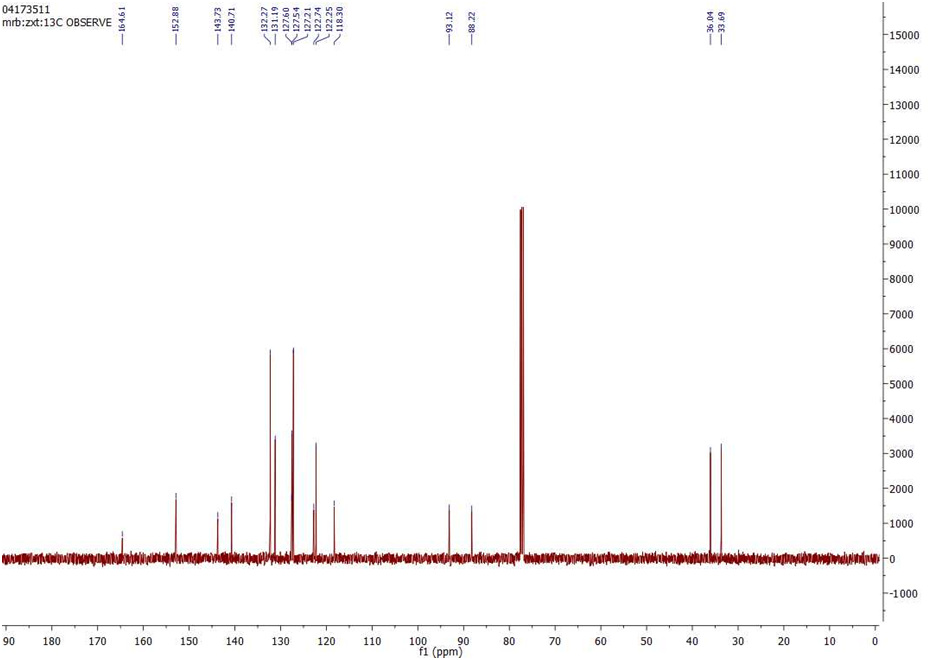}
\caption{\label{fig:s_NMR2}\ce{C^13} NMR Spectrum for $\mathbf{M}_\pi$.}
\end{figure}
\begin{figure}
\includegraphics[width=8.6cm]{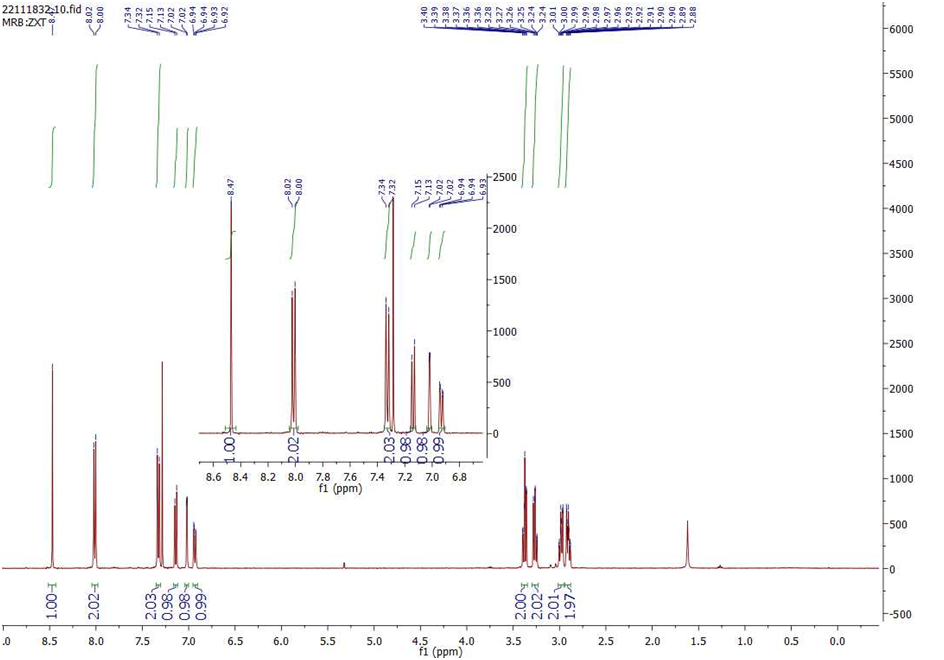}
\caption{\label{fig:s_NMR3}\ce{H^1} NMR Spectrum for $\mathbf{M}_\sigma$.}
\end{figure}
\begin{figure}
\includegraphics[width=8.6cm]{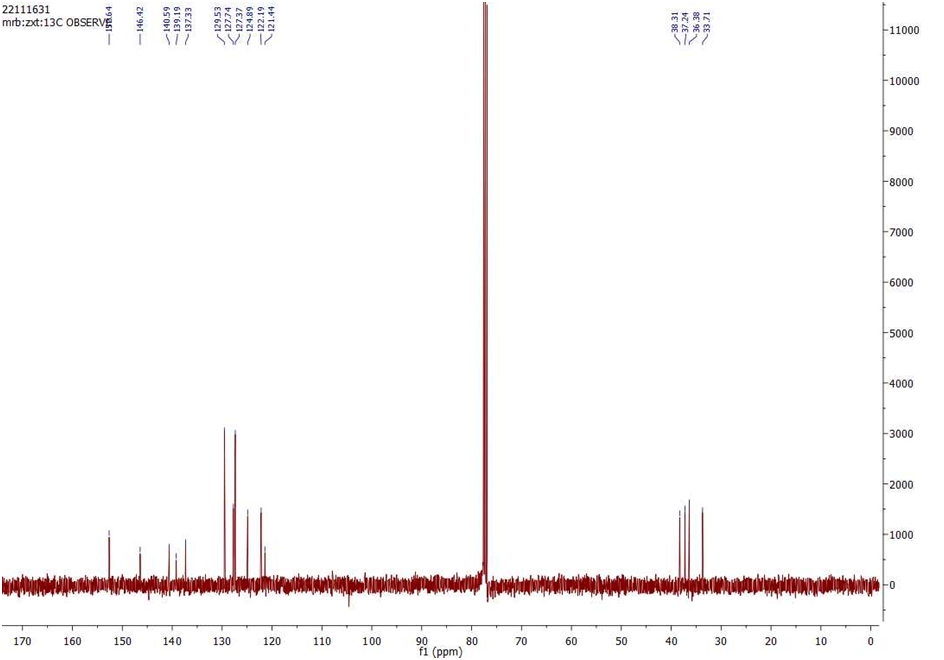}
\caption{\label{fig:s_NMR4}\ce{C^13} NMR Spectrum for $\mathbf{M}_\sigma$.}
\end{figure}

\section{Molecule solutions preparation}
Two stock solutions of $0.1~\textrm{M}$ molecule in 1:4 THF:TMB (v/v) were prepared. $0.1~\textrm{mM}$ solutions of $\mathbf{M}_\sigma$ and $\mathbf{M}_\pi$ were prepared from these. Three mixtures were also prepared from the stock solutions: 
\begin{enumerate}
\item \textbf{Mix 1} (1:3 \#/\#): $0.025~\textrm{mM}$ $\mathbf{M}_\sigma$ and $0.075~\textrm{mM}$ $\mathbf{M}_\pi$
\item \textbf{Mix 2} (1:1 \#/\#): $0.050~\textrm{mM}$ $\mathbf{M}_\sigma$ and $0.050~\textrm{mM}$ $\mathbf{M}_\pi$
\item \textbf{Mix 3} (3:1 \#/\#): $0.075~\textrm{mM}$ $\mathbf{M}_\sigma$ and $0.025~\textrm{mM}$ $\mathbf{M}_\pi$.
\end{enumerate}

\section{Experimental results}
\subsection{PC sorting on $\mathbf{X}_1$ using $PC_2$}
Although any PC (or combination of PCs) from the diagonalized correlation matrix [shown as an intensity plot in Fig. 2(b)] may be used to sort the data matrix ([Fig. 2(a)], to yield meaningful groups, the aim of this Letter was to demonstrate how PC sorting sorted $\mathbf{X}_1$ into groups based on which molecule,  $\mathbf{M}_\pi$ or $\mathbf{M}_\sigma$, was measured in the junction, and $PC_1$ successfully did this. 

\begin{figure}
\includegraphics[width=8.6cm]{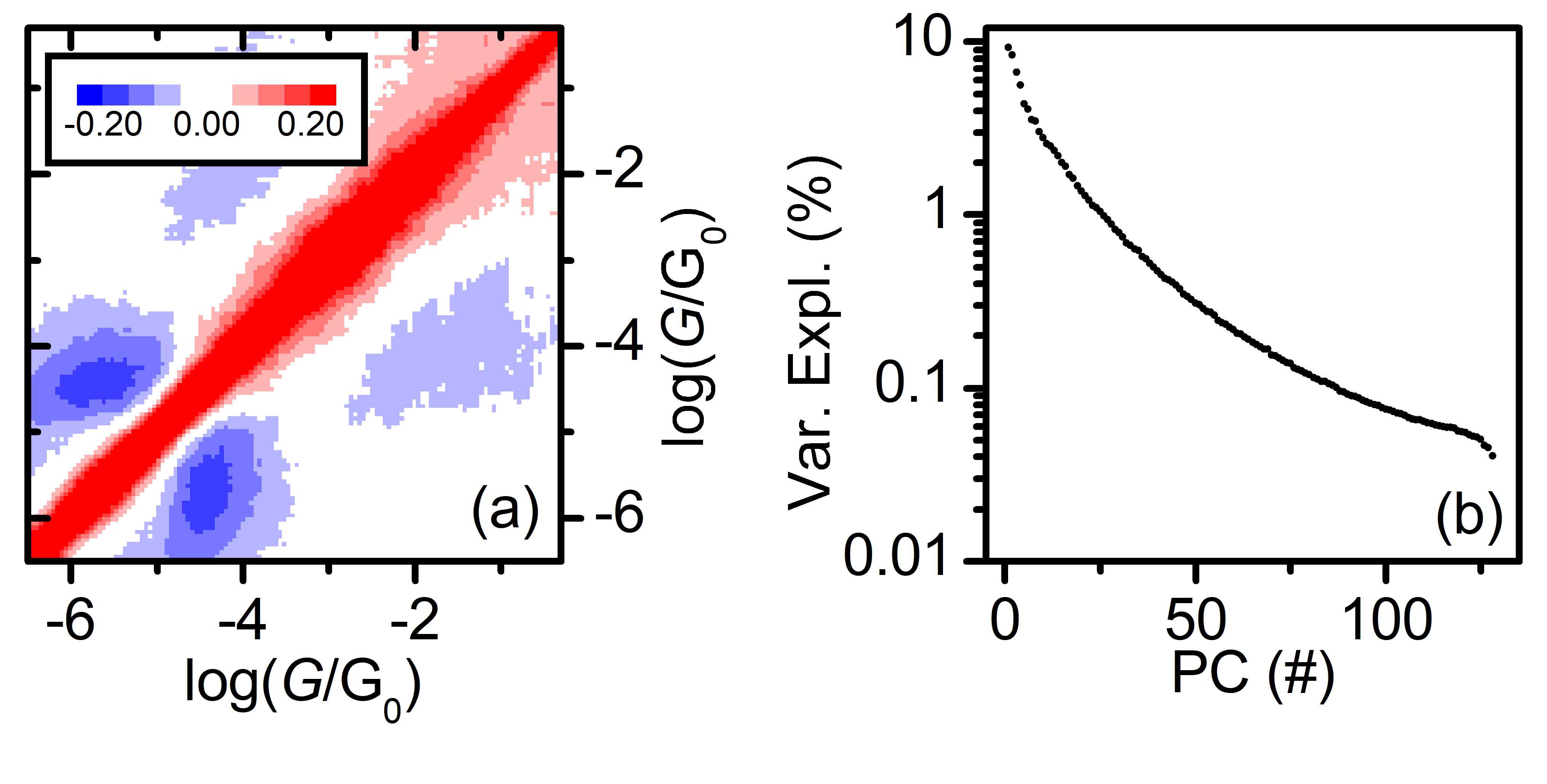}
\caption{\label{fig:s_seven} (a) Intensity plot of correlation matrix from $\mathbf{X}_1$, and (b) eigenvalues from correlation matrix as variance explained in a semilog plot.}
\end{figure}

\begin{figure}
\includegraphics[width=8.6cm]{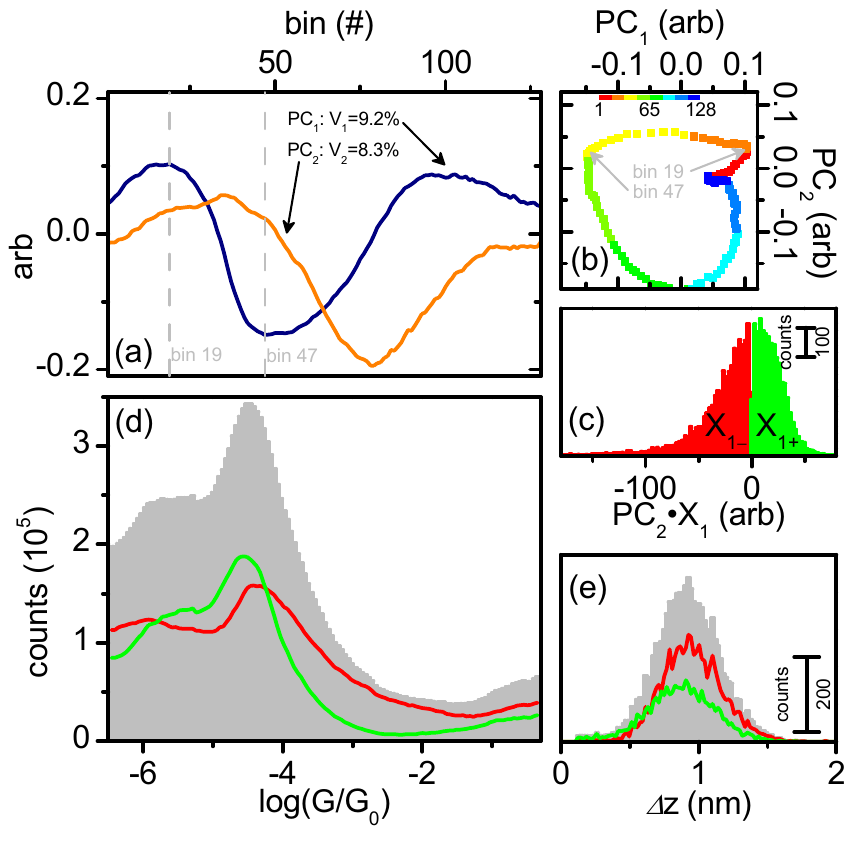}
\caption{\label{fig:s_eight} PCA analysis on \textbf{Mix 1} using $PC_2$.
(a) $PC_1$ (blue) and $PC_2$ (orange) plotted vs. bin number;
(b) $PC_1$ vs. $PC_2$ from bin 1 (red) to bin 128 (blue);
(c) histogram constructed from single 1D histograms projected onto $PC_2$ - curves in the red (green) bins had negative (positive) projections;
(d) negatively (7129 curves) and positively (4541 curves) projected curves were sorted into separate groups and total 1D histograms were constructed - gray histogram is total 1D histogram from entire data set (80\% of 14543 curves);
(e) plateau lengths were determined for each trace in the negative (red) and positive (green) groups and the entire data set (gray) and histograms were constructed for each.}
\end{figure}

For completeness, the results when $\mathbf{X}_1$ was sorted using $PC_2$ was shown here in Fig.~\ref{fig:s_eight}. Figure~\ref{fig:s_seven}(a) is a reproduction of Fig. 2(b) from the main text and Fig.~\ref{fig:s_seven}(b) is the eigenvalues, converted to variance explained, resulting from the diagonalization of the correlation matrix. Figures~\ref{fig:s_eight}(a,b) were identical to Figs. 3(a,b), but projecting $\mathbf{X}_1$ onto $PC_2$ instead of $PC_1$ yielded new results in Figs.~\ref{fig:s_eight}(c-e). Specifically, it was evident in Fig.~\ref{fig:s_eight}(d) that $PC_2$ had not sorted $\mathbf{X}_1$ into groups containing only $\mathbf{M}_\pi$ or $\mathbf{M}_\sigma$ junctions because both $\mathbf{X}_{1+}$ and $\mathbf{X}_{1-}$ did not have single mode distributions in their total 1DHs [Fig.~\ref{fig:s_eight}(d)]. $PC_2$ was largest in the bins corresponding to conductances of approximately $10^{-1}$ to $10^{-3}~\textrm{G}_0$, higher than either molecule. Figure~\ref{fig:s_seven}(a) contains a region of weak anticorrelation at approximately $(10^{-2}~\textrm{G}_0, 10^{-4}~\textrm{G}_0)$ which $PC_2$ may be describing. Instead of separating junctions based on criteria relating to the presence or absence of a molecule, or the presence of $\mathbf{M}_\pi$ or $\mathbf{M}_\sigma$, it was likely that sorting with $PC_2$ separated $\mathbf{X}_1$ into junctions which have a nearly ideal snap-back region, where the conductance trace drops quickly in the range $10^{-1}$ to $10^{-3}~\textrm{G}_0$ with no appreciable change in displacement, and junctions which exhibit some significant displacement in this same region. The difference in these two types of junctions can be seen in the 1D histograms with the presence or absence of counts in the bins corresponding to this conductance range. This presence or absence of counts was likely the variance in the data set which resulted in the large values in $PC_2$ in this conductance range. The likely causes of the non-ideal behavior in the snap-back region of the conductance traces were contamination or loss of mechanical stability.

\begin{figure}
\includegraphics[width=8.6cm]{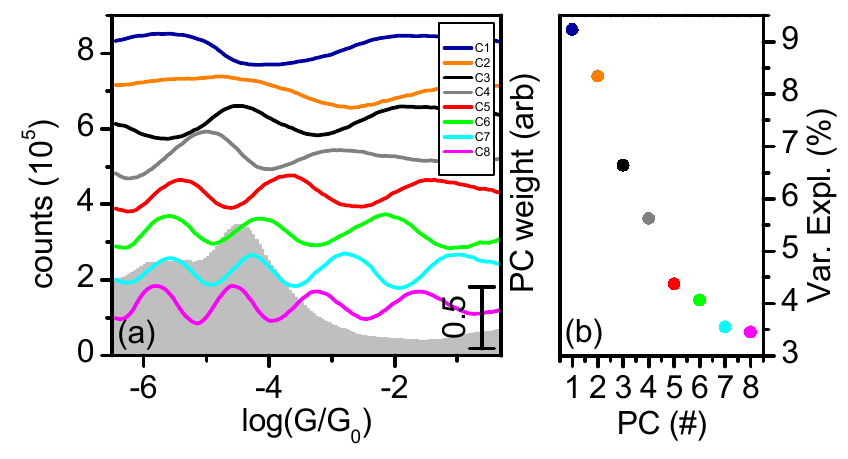}
\caption{\label{fig:s_nine} (a) left axis is the total 1D histogram (in grey) from $\mathbf{X}_1$ ; right axis is first 8 PCs from $\mathbf{X}_1$, and (b) eigenvalues plotted as variance explained in a linear plot.}
\end{figure}

Figure~\ref{fig:s_nine}(a) shows all eight of the first PCs from $\mathbf{X}_1$. The first 2 PCs (these were the two we used for analysis above and in the main text) had a qualitatively different nature than the remaining PCs. There appeared to be an emerging pattern as we looked at PCs $3-8$. At first glance PCs 3-8 had a sinusoidal structure, with each further PC having one more node than the previous. We did not wish to speculate on the nature of this pattern, except to suggest that whatever the cause, it was not highly descriptive of the molecular features. 

Figure~\ref{fig:s_nine}(b) plotted the first 8 eigenvalues, plotted as variance explained, on a linear scale. The semilog plot in Fig.~\ref{fig:s_seven}(b) already emphasized the extreme fall off of variance explained as we look at further PCs, and Fig.~\ref{fig:s_nine}(b) supported this claim.

To reiterate, many different PCs may be considered when applying this sorting method. However, we note that the rapid fall off of variance explained, as shown in Figs.~\ref{fig:s_seven}(b) and \ref{fig:s_nine}(b), suggests that this specific experiment contains much of the salient variance in the first few PCs. We also note that the emergence of a regular pattern in PCs $3-8$, shown in Fig.~\ref{fig:s_nine}(a), indicates that those PCs are governed by variances not involving the molecules. Finally, we emphasize our stated goal: to find a simple and fast approach to meaningfully separate two groups from a single data set. If we find a PC which does this in the first or second PC, then that PC will explain larger variance than any subsequent PC we may find.

\subsection{PC sorting on two further mixtures of $\mathbf{M}_\pi$ and $\mathbf{M}_\sigma$: \textbf{Mix 2} and \textbf{Mix 3}}
\begin{figure}
\includegraphics[width=8.6cm]{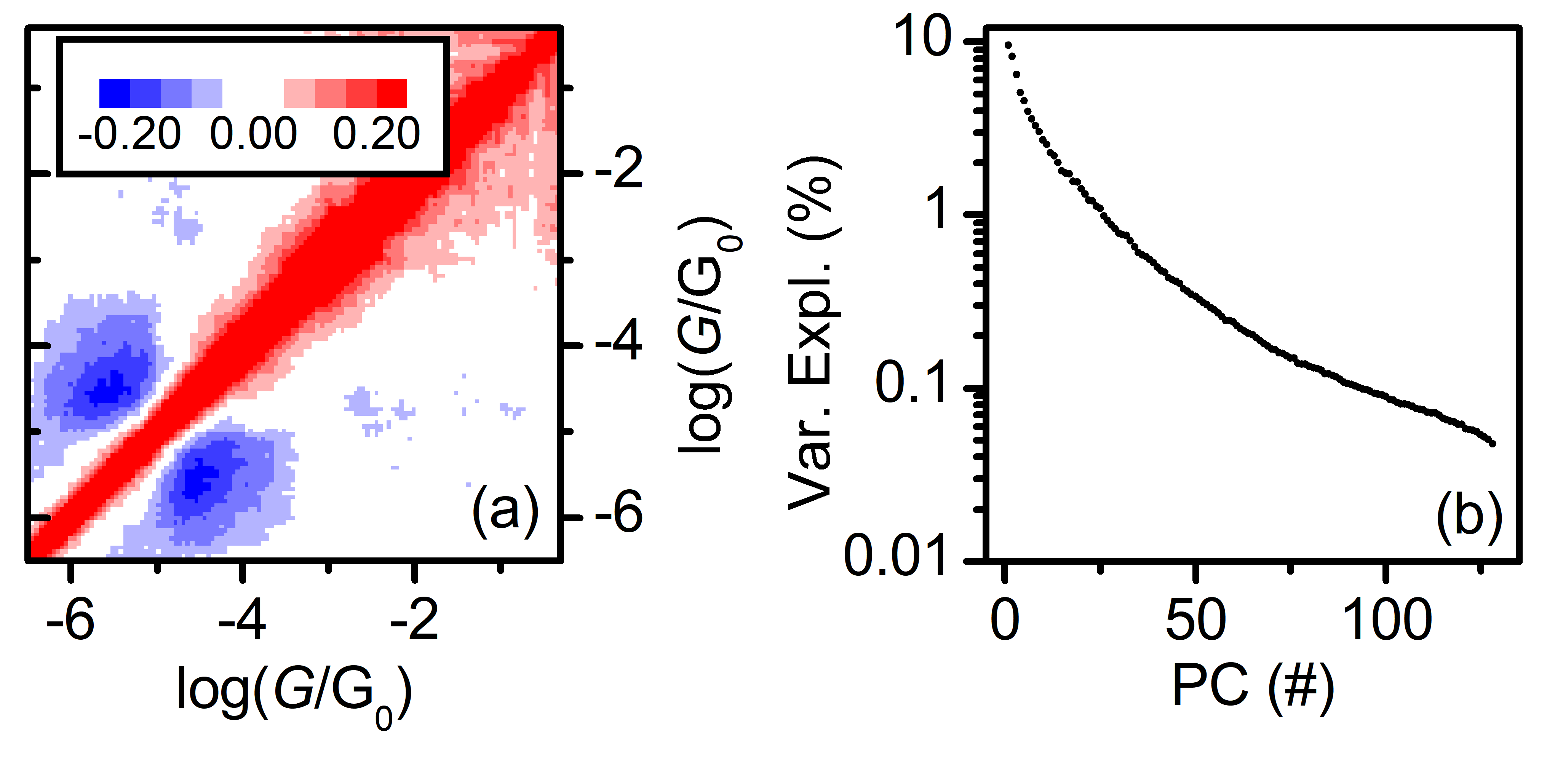}
\caption{\label{fig:s_ten} (a) Intensity plot of correlation matrix from $\mathbf{X}_2$, and (b) eigenvalues from correlation matrix as variance explained in a semi-log plot.}
\end{figure}

\begin{figure}
\includegraphics[width=8.6cm]{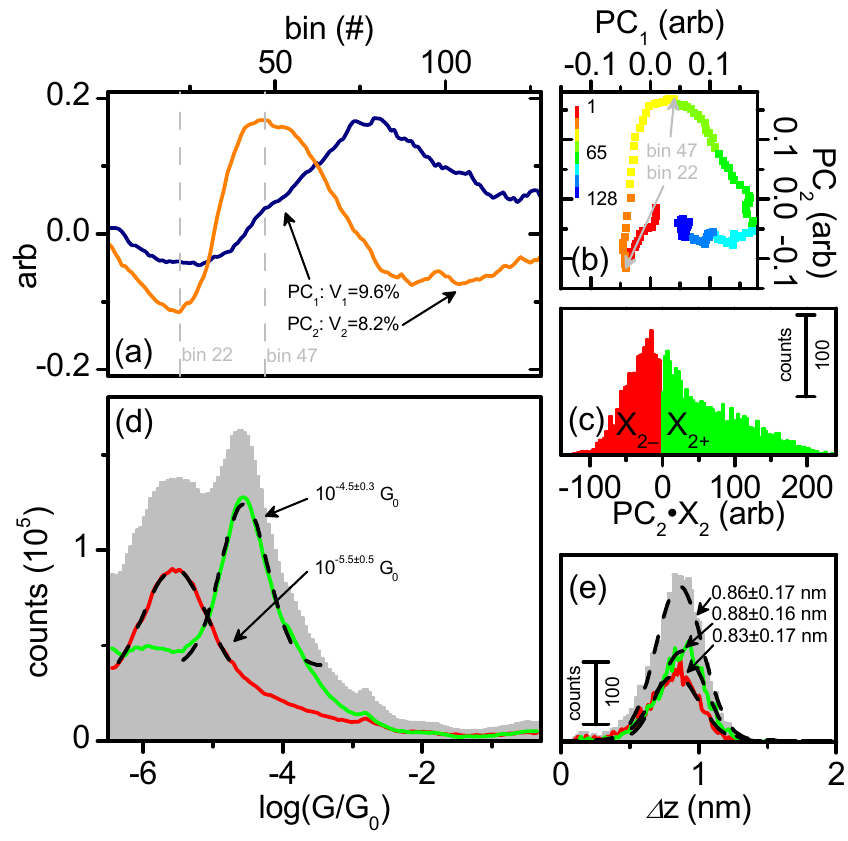}
\caption{\label{fig:s_eleven} PC sorting on \textbf{Mix 2} using $PC_2$.
(a) $PC_1$ (blue) and $PC_2$ (orange) plotted vs. bin number;
(b) $PC_1$ vs. $PC_2$ from bin 1 (red) to bin 128 (blue);
(c) histogram constructed from single 1D histograms projected onto $PC_2$ - curves in the red (green) bins had negative (positive) projections;
(d) negatively (7129 curves) and positively (4541 curves) projected curves were sorted into separate groups and total 1D histograms were constructed - gray histogram is total 1D histogram from entire data set (80\% of 14543 curves);
(e) plateau lengths were determined for each trace in the negative (red) and positive (green) groups and the entire data set (gray) and histograms were constructed for each.}
\end{figure}

\begin{figure}
\includegraphics[width=8.6cm]{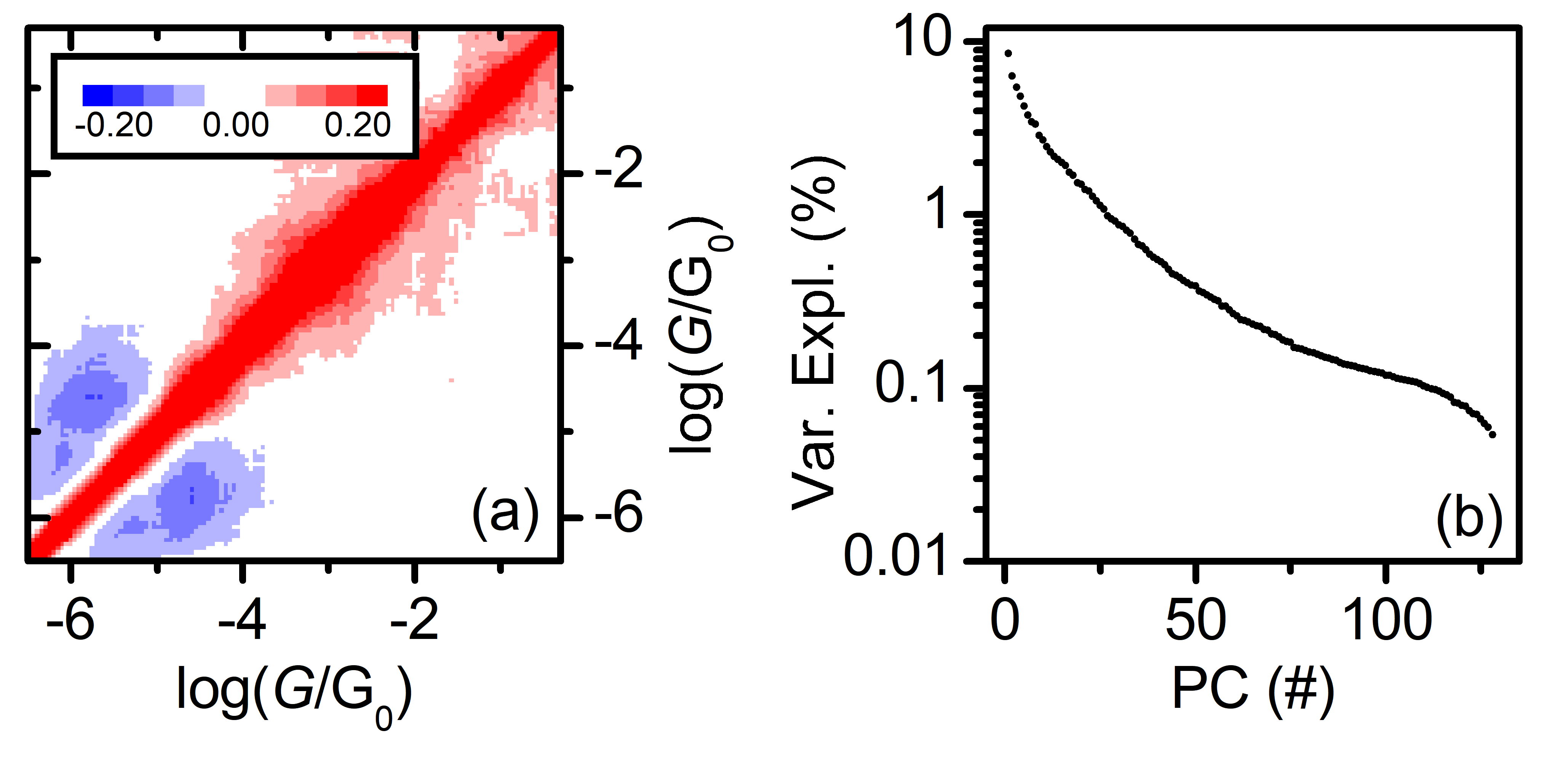}
\caption{\label{fig:s_twelve} (a) Intensity plot of correlation matrix from $\mathbf{X}_3$, and (b) eigenvalues from correlation matrix as variance explained in a semi-log plot.}
\end{figure}

\begin{figure}
\includegraphics[width=8.6cm]{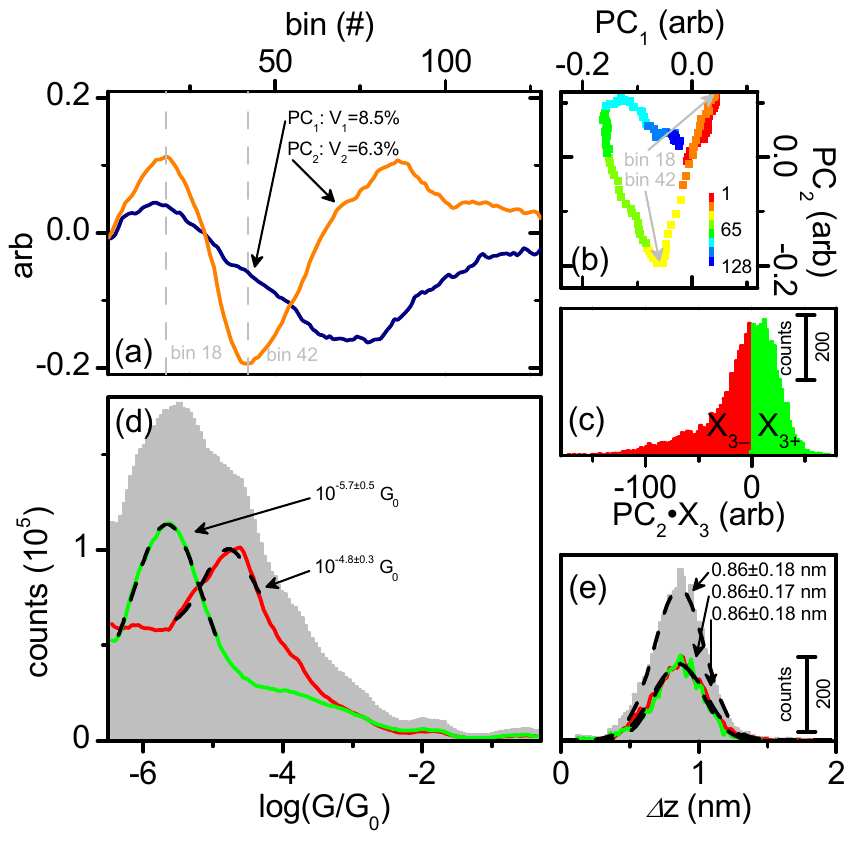}
\caption{\label{fig:s_thirteen} PC sorting on \textbf{Mix 3} using $PC_2$.
(a) $PC_1$ (blue) and $PC_2$ (orange) plotted vs. bin number;
(b) $PC_1$ vs. $PC_2$ from bin 1 (red) to bin 128 (blue);
(c) histogram constructed from single 1D histograms projected onto $PC_2$ - curves in the red (green) bins had negative (positive) projections;
(d) negatively (7129 curves) and positively (4541 curves) projected curves were sorted into separate groups and total 1D histograms were constructed - gray histogram is total 1D histogram from entire data set (80\% of 14543 curves);
(e) plateau lengths were determined for each trace in the negative (red) and positive (green) groups and the entire data set (gray) and histograms were constructed for each.}
\end{figure}

To demonstrate further the effectiveness of PC sorting we measured two more systems. Besides the 1:3 $\mathbf{M_\sigma}$:$\mathbf{M_\pi}$ mixture, \textbf{Mix 1}, described in the main text, and above, two further mixtures, \textbf{Mix 2} and \textbf{Mix 3} were prepared and measured with MCBJ. Correlation matrices, plotted as intensity plots, can be seen in Figs.~\ref{fig:s_ten} and \ref{fig:s_twelve}(a)and the eigenvalues, converted to variance explained, can be seen in Figs.~\ref{fig:s_ten} and \ref{fig:s_twelve}(b). For both \textbf{Mix 2} and \textbf{Mix 3}, $PC_2$ was most successful at sorting the junctions into groups of $\mathbf{M}_\pi$ and $\mathbf{M}_\sigma$ junctions. In both cases, $PC_2$ [Figs.~\ref{fig:s_eleven},\ref{fig:s_thirteen}(a)] had maxima and minima in the same bins corresponding to molecular conductances of $\mathbf{M}_\pi$ and $\mathbf{M}_\sigma$, and thus was the logical choice. Plots of $PC_1$ vs $PC_2$ [Figs.~\ref{fig:s_eleven} and \ref{fig:s_thirteen}(b)] clarified this. 

Using PC sorting, $\mathbf{X}_2$ was sorted [Fig.~\ref{fig:s_eleven}(c)] into 2505 curves in $\mathbf{X}_{2-}$ and 3196 curves in $\mathbf{X}_{2+}$. $\mathbf{X}_{2-}$ and $\mathbf{X}_{2+}$ were associated with junctions involving $\mathbf{M}_\sigma$ and $\mathbf{M}_\pi$, respectively, by comparing the 1DHs [Fig.~\ref{fig:s_eleven}(d)] with $\mathbf{M}_\sigma$ and $\mathbf{M}_\pi$ measured separately. The average plateaus were the same length [Fig.~\ref{fig:s_eleven}(e)].

PC sorting of $\mathbf{X}_3$ garnered strikingly similar results, except the numbers of junctions involving $\mathbf{M}_\sigma$ and $\mathbf{M}_\pi$ changed (5265 $\mathbf{M}_\pi$ and 4823 $\mathbf{M}_\sigma$). Also, in the case of $\mathbf{X}_3$, $\mathbf{X}_{3-}$ was associated with junctions involving $\mathbf{M}_\pi$, while $\mathbf{X}_{3+}$ was associated with $\mathbf{M}_\sigma$.

\subsection{PC sorting on $\mathbf{X}_\pi$ using $PC_2$}
\begin{figure}
\includegraphics[width=8.6cm]{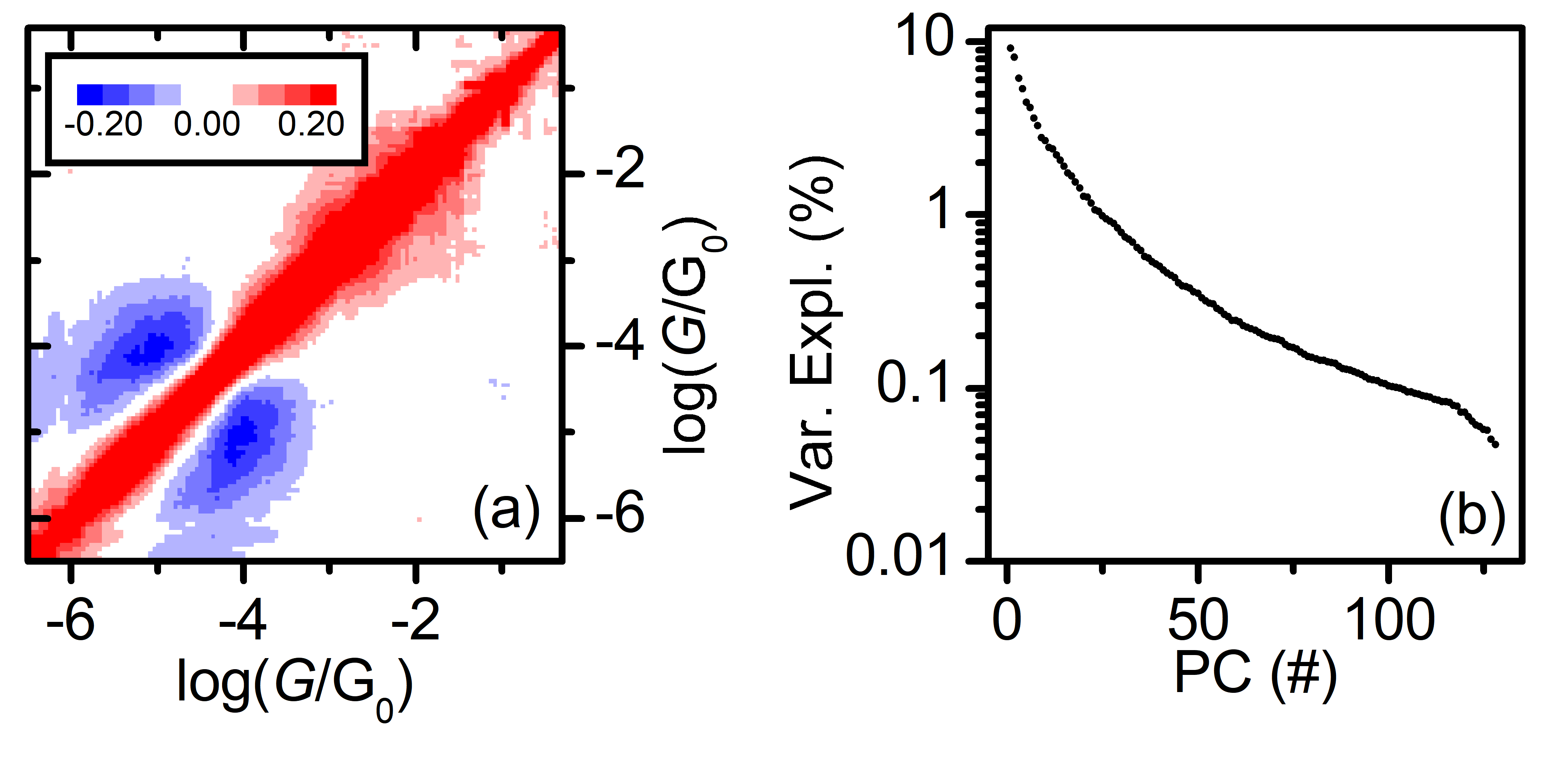}
\caption{\label{fig:s_fourteen} (a) Intensity plot of correlation matrix from $\mathbf{X}_\pi$, and (b) eigenvalues from correlation matrix as variance explained in a semi-log plot.}
\end{figure}

\begin{figure}
\includegraphics[width=8.6cm]{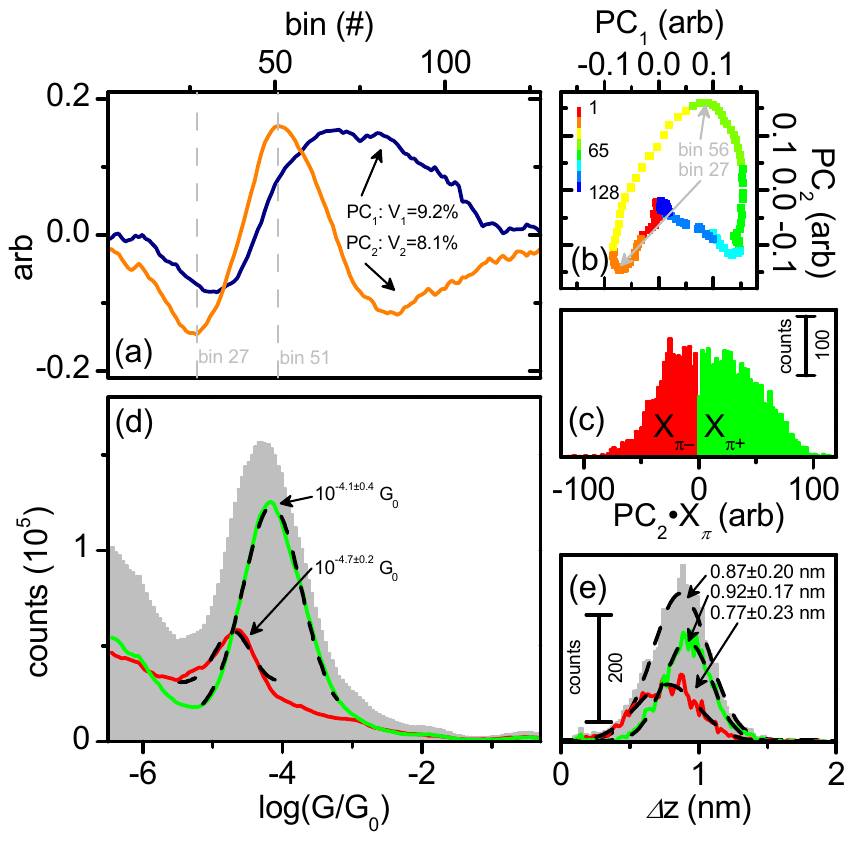}
\caption{\label{fig:s_fifteen} PC sorting on $\mathbf{X}_\pi$ using $PC_2$.
(a) $PC_1$ (blue) and $PC_2$ (orange) plotted vs. bin number;
(b) $PC_1$ vs. $PC_2$ from bin 1 (red) to bin 128 (blue);
(c) histogram constructed from single 1D histograms projected onto $PC_2$ - curves in the red (green) bins had negative (positive) projections;
(d) negatively (7129 curves) and positively (4541 curves) projected curves were sorted into separate groups and total 1D histograms were constructed - gray histogram is total 1D histogram from entire data set (80\% of 14543 curves);
(e) plateau lengths were determined for each trace in the negative (red) and positive (green) groups and the entire data set (gray) and histograms were constructed for each.}
\end{figure}

The PC sorting steps which yielded Fig. 4 are shown here. Diagonalizing the correlation matrix [intensity plot in Fig.~\ref{fig:s_fourteen}(a)] yielded eigenvalues, converted to variance explained, plotted in Fig.~\ref{fig:s_fourteen}(b). $PC_1$ and $PC_2$ [Figs.~\ref{fig:s_fifteen}(a,b)] were very similar in the bins involving the molecular conductance of $\mathbf{M}_\pi$, but $PC_2$ was slightly more dominant in the pertinent conductance range, so it was used to sort $\mathbf{X}_\pi$ into $\mathbf{X}_{\pi-}$ and $\mathbf{X}_{\pi+}$ [Fig.~\ref{fig:s_fifteen}(c)].

\subsection{Master curve fitting}
\begin{figure}
\includegraphics[width=8.6cm]{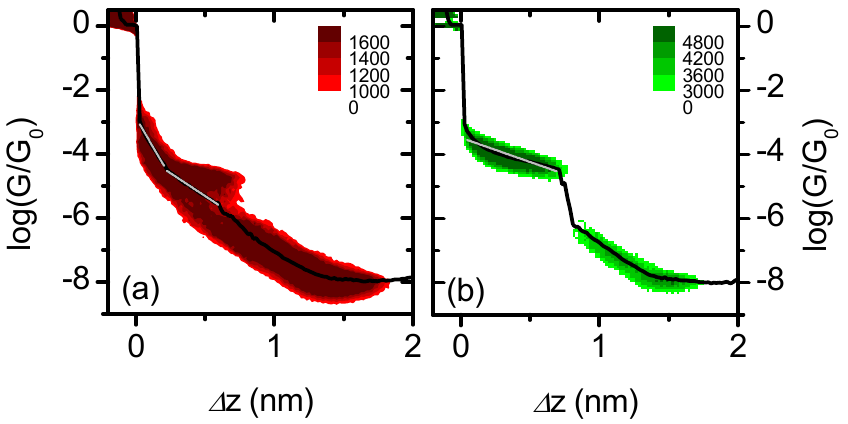}
\caption{\label{fig:s_sixteen} Master curves fit to 2D histograms from (a) $\mathbf{X}_{\pi-}$ and (b) $\mathbf{X}_{\pi+}$. Grey lines are linear fits to molecular plateaus. Master curve in (a) was fit to two lines with two separate slopes.}
\end{figure}

The vertical bins in the 2D histogram (displacement) can each be fit to a Gaussian following procedures previously published \cite{Moreno-Garcia2013}. The mean of each fit can be assembled into an array, called the master curve, which describes the average trace in the data set. 2D histograms for $\mathbf{X}_{\pi-}$ and $\mathbf{X}_{\pi+}$ were fit to determine master curves [Figs.~\ref{fig:s_sixteen}(a,b)]. The molecular plateau region was then linearly fit to determine the slope of the plateau. In the case of $\mathbf{X}_{\pi-}$, two regions were fit to two separate slopes. The slopes were converted from $log_{10}$ to $log_{e}$ for purposes of referring to Eq. 1 (in the main text) and Ref.~\cite{Huang2015break}. 

\bibliographystyle{apsrev4-1}

%